\documentclass[comment, prl,twocolumn,nofootinbib, preprintnumbers, superscriptaddress]{revtex4}

\usepackage{amsmath,amssymb,bm,slashed,braket}
\usepackage{graphicx}
\usepackage{epstopdf}
\usepackage{float}
\usepackage[colorlinks=true,
            linkcolor=blue,
            urlcolor=blue,
            citecolor=green,          
            bookmarks=true,
            bookmarksnumbered=true,
            breaklinks=true,
            pdfpagemode=Fullscreen,
            pdfstartview=FitBH]{hyperref}

\usepackage[normalem]{ulem}
\usepackage{color}

\definecolor{Orange}{cmyk}{0,0.61,0.87,0}
\definecolor{JungleGreen}{cmyk}{0.99,0,0.52,0}
\definecolor{OliveGreen}{cmyk}{0.64,0,0.95,0.40}
\definecolor{Brown}{cmyk}{0,0.81,1,0.60}
\definecolor{RoyalBlue}{cmyk}{0.71,0.53,0,0.12}
\definecolor{Gray}{cmyk}{0,0,0,0.40}
\definecolor{LightPink}{cmyk}{0.0,0.25,0,0}
\definecolor{LLightPink}{cmyk}{0.0,0.10,0,0}
\definecolor{LightBlue}{cmyk}{0.25,0,0,0}
\definecolor{LightGray}{cmyk}{0,0,0,0.2}

\usepackage{xcolor}
\definecolor{gesfpurple}{rgb}{0.47,0.19,0.42}

\definecolor{gesflanse}{rgb}{0.00,0.50,0.50}

\definecolor{gesfblue}{rgb}{0.08,0.42,0.76}

\definecolor{gesfred}{rgb}{1,0,0}

\definecolor{gesfwhite}{rgb}{1,1,1}

\definecolor{gesfblack}{rgb}{0,0,0}

\newcommand{\geqn}[1]{Eq.\,\hypersetup{linkcolor=blue}(\ref{#1})\hypersetup{linkcolor=blue}}
\newcommand{\gfig}[1]{{\hypersetup{linkcolor=violet}Fig.\,\ref{#1}\hypersetup{linkcolor=blue}}}
\newcommand{\gtab}[1]{{\hypersetup{linkcolor=gesflanse}Table\,\ref{#1}\hypersetup{linkcolor=blue}}}

\graphicspath{{figs/}}

\begin{document}

\title{
Dark Photon Searches with Initial-State Radiation at Fixed-Target Configurations \\
}

\author{Shao-Feng Ge}
\email{gesf@sjtu.edu.cn}
\affiliation{State Key Laboratory of Dark Matter Physics, Tsung-Dao Lee Institute \& School of Physics and Astronomy, Shanghai Jiao Tong University, China}
\affiliation{Key Laboratory for Particle Astrophysics and Cosmology (MOE) \& Shanghai Key Laboratory for Particle Physics and Cosmology, Shanghai Jiao Tong University, Shanghai 200240, China}

\author{Jinhan Liang}
\email{jinhanliang@m.scnu.edu.cn}
\affiliation{State Key Laboratory of Nuclear Physics and Technology, Institute of Quantum Matter, South China Normal University, Guangzhou 510006, China}
\affiliation{Guangdong Basic Research Center of Excellence for Structure and Fundamental Interactions of Matter, Guangdong Provincial Key Laboratory of Nuclear Science, Guangzhou
510006, China}

\author{Zuowei Liu}
\email{zuoweiliu@nju.edu.cn}
\affiliation{Department of Physics, Nanjing University, Nanjing 210093, China}

\author{Ui Min}
\email{ui.min@sjtu.edu.cn}
\affiliation{State Key Laboratory of Dark Matter Physics, Tsung-Dao Lee Institute \& School of Physics and Astronomy, Shanghai Jiao Tong University, China}
\affiliation{Key Laboratory for Particle Astrophysics and Cosmology (MOE) \& Shanghai Key Laboratory for Particle Physics and Cosmology, Shanghai Jiao Tong University, Shanghai 200240, China}

\begin{abstract}
In this work, we investigate
the contribution of the annihilation process with initial-state radiation
($e^+ e^- \to \gamma A'$)
to the invisible dark photon ($A'$) searches at the electron
fixed-target configurations. For illustration, we
consider both the disappearing positron track signature
at Belle II and the large missing energy search at
NA64.
When the dark photon has a narrow decay width,
the effect of the initial-state radiation to the annihilation process
can dominate over its $s$-channel and bremsstrahlung counterparts 
around
$m_{A'} \simeq 60\,\rm{MeV}$
($m_{A'} \simeq 200\,\rm{MeV}$) for Belle II (NA64),
to enhance the corresponding sensitivity on the
kinetic mixing parameter $\epsilon$ by a factor of
up to approximately 2.7 (1.3).
For Belle II, we further perform a multi-bin analysis
with the spectrum information to better separate
the background and signal channels for significant
improvement of the sensitivity.
\end{abstract}

\maketitle

\section{\bf Introduction}

Although dark matter (DM) makes up about 
a quarter of the total energy density of the Universe, 
its particle properties remain unknown
\cite{Bertone:2004pz,Feng:2010gw,Young:2016ala,Cooley:2022ufh,Cirelli:2024ssz}. 
A particularly intriguing class of DM models 
involves couplings to the 
Standard Model (SM) particles 
via portal interactions, 
with a notable example of
the dark photon (vector) portal \cite{Fabbrichesi:2020wbt}.  
The dark photon 
can arise from either
the kinetic mixing \cite{Holdom:1985ag,Foot:1991kb} 
or the Stueckelberg mass mixing 
\cite{Kors:2005uz, Feldman:2006ce, Feldman:2006wb, Cheung:2007ut, Feldman:2007wj} 
between the SM hypercharge boson and
a new gauge boson in the hidden sector.  
A wide range of experimental strategies have been
developed to search for the dark photon, including cosmological and astrophysical probes, 
precision electromagnetic measurements, 
and accelerator-based experiments
\cite{Jaeckel:2010ni,Fabbrichesi:2020wbt}.

The dark photon in the MeV-GeV mass range that
predominantly decays into dark sector particles
can be most effectively probed by
accelerator-based experiments, including BaBar \cite{BaBar:2008aby, BaBar:2017tiz}, and Belle II \cite{Belle-II:2018jsg, Liang:2022pul}, and NA64 \cite{Banerjee:2019pds, Andreev:2021fzd, NA64:2023ehh}.

The NA64 experiment can operate in both $e^-$ and $e^+$ modes, using an electron or positron beam impinging on an active target (calorimeter), and is highly sensitive to dark photons via the missing-energy signature.
The dark photon can be produced either via the electron bremsstrahlung process on a nuclear target \cite{Andreas:2013lya} or through $e^+ e^-$ annihilation, where the electron originates from the target atom
and the positron is generated in the electromagnetic shower of the electron/positron beams through their track-length distributions
\cite{Marsicano:2018glj, Andreev:2021fzd, NA64:2023ehh}.

The annihilation process enables NA64 to place the most stringent 
constraint on the dark photon with mass $\sim$300\,MeV,
which corresponds to the center-of-mass energy
of the $e^+e^-$ annihilation process 
\cite{Marsicano:2018glj, Andreev:2021fzd}. 
Belle II and BaBar are electron-positron colliders 
operating at a center-of-mass energy of $\sim$10\,GeV, 
and can be used to 
search for the invisibly decaying dark photon via 
the mono-photon channel.
In this case, dark photons are produced 
through the initial-state radiation (ISR) process,
$e^+ e^- \to \gamma A'$ \cite{BaBar:2017tiz,Belle-II:2018jsg}. 
The radiative photon shifts the invariant mass of the
initial electron-positron system such that the dark
photon $A'$ can be {\it resonantly} produced at
electron colliders, which is the so-called radiative
return process \cite{Denig:2006kj,Karliner:2015tga}.

Recently, Ref.~\cite{Liang:2022pul} proposed using Belle II 
in a fixed-target mode to search for dark photons, 
with the final-state positron of the primary
Bhabha scattering $e^+e^- \to e^+e^-$ 
serving as the incident particle
and the calorimeter as the target.\footnote{
Although Belle II is an accelerator experiment,
the final-state positron is essentially dumped
in the detector which serves as a fixed target.
For this reason, we use {\it fixed-target configuration} rather
than {\it fixed-target experiment} for both NA64 and Belle II.}
This detection channel has been dubbed as the 
``{\it disappearing positron track}'' signature
at Belle II, 
and was shown to provide leading sensitivity to
the dark photon
with mass $\sim$70\,MeV. 
However, the previous work 
only considered dark photon production via 
the electron bremsstrahlung  
and the $s$-channel process ($e^+e^- \to A'$). 

In the current study, we extend the previous dark photon
searches at both Belle II and NA64 by including the
ISR effect and discuss
the $\alpha_D$ dependence on the experimental
sensitivities, where $\alpha_D \equiv g_D^2/(4\pi)$,
with $g_D$ denoting the coupling constant between
the dark photon and dark fermions. 
We find that ISR distorts the shape of the 
dark photon resonance and can significantly strengthen the constraints in its certain 
mass regions, particularly when the 
dark photon has a narrow decay width.
Since the ISR effect allows resonant production of
dark photon with a smaller mass, a significant sensitivity enhancement is achieved for the 
dark photon with
mass below 65\,MeV at Belle II and around 200\,MeV
at NA64.
Comparing with the single-bin analysis in \cite{Liang:2022pul},
the high energy resolution of the Belle II calorimeter allows us 
to carry out a binned analysis of the disappearing
positron track signature, which can further
strengthen the sensitivity by up to 30\%.

\section{\bf Dark Photon Production with Initial-State Radiation}

Connecting the visible and dark sectors with
dark photon is a very economical scheme.
The Lagrangian for dark photon models with a dark 
fermion $\chi$ charged under $U(1)_X$ is given by
\cite{Fabbrichesi:2020wbt,Holdom:1985ag,Foot:1991kb,Kors:2005uz,Feldman:2006ce,Feldman:2006wb,Cheung:2007ut,Feldman:2007wj,Jaeckel:2010ni}
\begin{align}
  \mathcal{L}_{\rm DP}
=
- \frac{1}{4} F'_{\mu \nu} F'^{\mu \nu}
+ \frac{\epsilon}{2} F'_{\mu \nu} F^{\mu \nu}
+ g_D A'_\mu \bar{\chi} \gamma^\mu \chi,
\end{align}
where $\epsilon \ll 1$ is the kinetic mixing parameter between
the SM photon $A$ and the dark photon $A'$. The dark gauge
coupling strength $g_D$ is from the dark $U(1)_X$
gauge group. We assume that the dark photon acquires
its mass $m_{A'}$ via spontaneous breaking of
the $U(1)_X$ symmetry or via the Stueckelberg
mechanism. After canonical normalization of the photon
and dark photon fields, $A_\mu\rightarrow A_\mu +
\epsilon A'_\mu + \mathcal{O}(\epsilon^2)$
and $A'_\mu \rightarrow A'_\mu+\mathcal{O}
(\epsilon^2)$, the dark photon interacts with the
SM fermions $f_{\rm SM}$ as,
\begin{align}
  \epsilon Q_f e  A'_\mu \bar{f}_{\rm SM} \gamma^\mu f_{\rm SM} ~,
\end{align}
where $Q_f$ is the 
electric charge of $f_{\rm SM}$.

For fixed-target configurations with incident electrons
or positrons, the dark photon can be produced through
the annihilation and bremsstrahlung processes \cite{Marsicano:2018glj}.
Bremsstrahlung occurs when the incoming electron
or positron scatters off a target nucleus, with a
cross section that increases as the dark photon mass
decreases and the corresponding cross section formulas
can be found in \cite{Bjorken:2009mm, Gninenko:2017yus, Liu:2017htz}.
The annihilation processes arise from
the annihilation either of the incident positron or  
of the secondary positron produced by the electromagnetic
shower induced by the incoming electron, with the
atomic electron in the target.
The annihilation process can contribute
significantly when the 
invariant mass of the positron and an
atomic electron is close to the dark photon mass.

Without considering the ISR effect, the 
cross section
of the $e^+ e^- \to A'^{(*)} \to \chi \bar{\chi}$
annihilation process at the leading order is
\begin{subequations}
\begin{align}
&
\sigma_{\rm ann}^0
=
\frac{4\pi}{3}\epsilon^2 \alpha_e \alpha_D \frac{(s_{\chi\bar{\chi}} + 2 m^2_\chi)\sqrt{1- \frac {4 m^2_\chi}{s_{\chi\bar{\chi}}}}}{(s_{\chi\bar{\chi}} - m^2_{A'})^2 + m^2_{A'} \Gamma^2_{A' \to \chi \bar{\chi}} } ~,
\\ &
\Gamma_{A' \to \chi \bar{\chi}} = \frac{1}{3} \alpha_D m_{A'} \left(1 + \frac{2m^2_\chi}{m^2_{A'}} \right) \sqrt{1 - \frac{4m^2_\chi}{m^2_{A'}} } ~,
\label{eq:xsec-s-chan}
\end{align}
\end{subequations}
where $s_{\chi\bar{\chi}}$ denotes the squared
invariant mass of the final-state DM pair,
$m_\chi$ is the DM mass,
$\Gamma_{A' \to \chi\bar{\chi}}$ is the 
dark photon invisible decay width, and $\alpha_D \equiv g_D^2/(4\pi)$.
For the parameter space of interest in this work, the invisible decay width dominates over the visible one.

\begin{figure}
\centering
\includegraphics[width=0.23\textwidth]{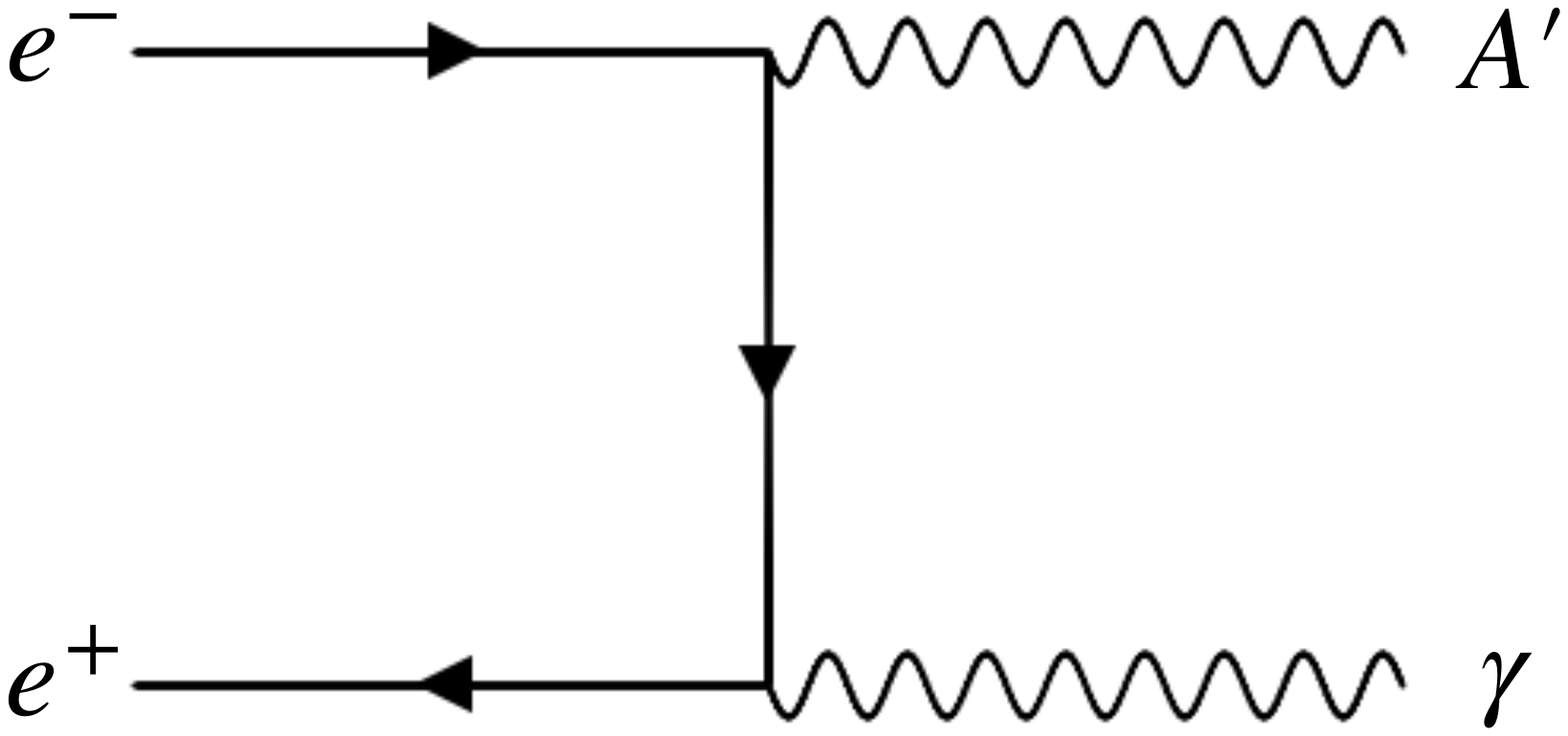} 
\hfill
\includegraphics[width=0.23\textwidth]{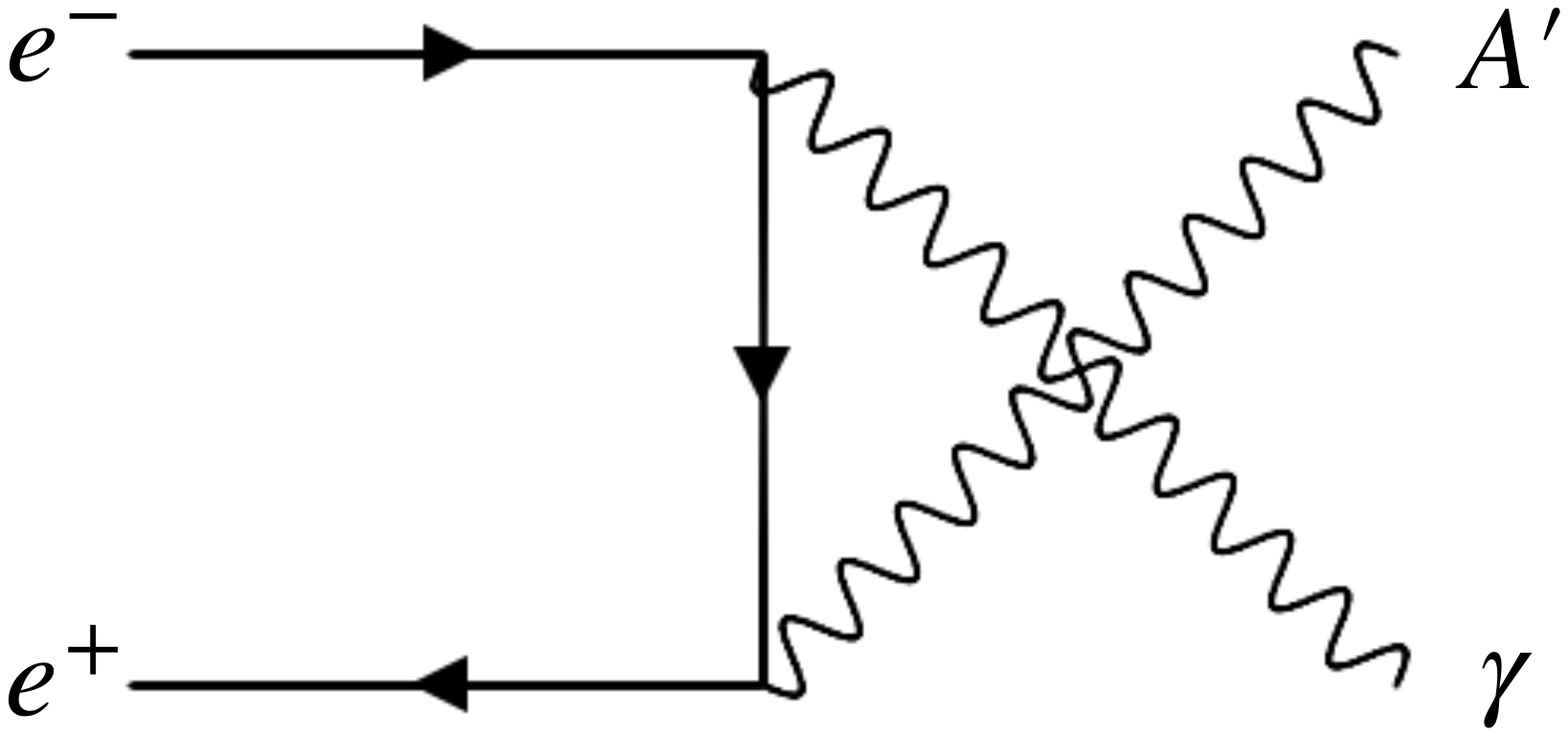} 
\caption{Feynman diagrams of $e^+e^- \rightarrow \gamma A'$
processes
considering the $s$-channel with an initial-state radiation
for the dark photon production.
}
\label{fig:Feynman_diagram}
\end{figure}

However, an extra photon can be produced besides the
dark photon due to the ISR effect, as shown in
\gfig{fig:Feynman_diagram}. 
For NA64 and for
the disappearing-positron-track signature at Belle II \cite{Liang:2022pul}, 
the ISR photon cannot be reconstructed, 
since its deposited energy is indistinguishable from that of the incoming particles. 
Consequently, 
annihilation events with and without ISR cannot be separated through a photon energy cut, 
in contrast with standard collider analyses.
Note that including the ISR effect
for the large missing energy signature
has not been considered in the dark photon searches
at NA64 \cite{Banerjee:2019pds, Andreev:2021fzd}.
Although it is mentioned in the theoretical study of
\cite{Marsicano:2018krp}, their analysis is limited to
the case of a dark photon with a large decay width
($\alpha_D = 0.5$), where the ISR contribution
is actually subdominant in comparison with the $s$-channel and
bremsstrahlung processes, as we elaborate in later
sections.

To properly include the contributions from ISR processes, 
we work within the electron PDF framework, 
where the virtual corrections are incorporated and  
cancel the infrared divergence associated with real ISR emission.
In practice, this requires incorporating the next-to-leading-order (NLO) diagrams.
We therefore compute the dark photon production rate using the radiator function given  
in \cite{Nicrosini:1988hw,Ping:2013jka}, which was derived in the electron-PDF formalism.

In the center-of-mass (CoM) frame, the fully
differential cross section with ISR included 
can be written as \cite{Nicrosini:1988hw,Ping:2013jka}
\begin{align}
\frac{d\sigma_{\rm ann}^{\rm NLO}}{dc_\gamma^* dx_\gamma^*}
= \mathcal{R}(x_\gamma^*, c_\gamma^*,s_{ee}) \sigma_{\rm ann}^0 (s_{\chi\bar{\chi}}),
\label{eq:exc}
\end{align}
where 
$s_{ee}$ denotes the squared invariant mass of the initial $e^+ e^-$ system.
We denote those CoM quantities with an asterisk ($*$),
including $c_\gamma^* \equiv \cos\theta_\gamma^*$ 
with $\theta_\gamma^*$ denoting the polar angle of the
ISR photon with respect to the positron momentum and 
$x_\gamma^* \equiv 2 E_\gamma^* / \sqrt{s_{ee}}$ 
with $E_\gamma^*$ being the photon energy.
In addition,
$s_{\chi\bar{\chi}} \equiv (1-x_\gamma^*)s_{ee}$ is for
the $\chi \bar \chi$ system 
and $\mathcal{R}$ is the radiator function. 
Note that $x_\gamma^*$ 
represents the energy fraction carried by the ISR photon from the beam particles.
In \geqn{eq:exc}, the NLO annihilation cross section $\sigma^{\rm NLO}_{\rm ann}$ accounts for the contributions from both real emissions at tree-level and virtual 
corrections at one-loop level.

The radiator $\mathcal{R}(x_\gamma^*, c_\gamma^*, s_{ee})$ represents the probability for an electron or positron to emit a photon with an energy fraction $x_\gamma^*$ and polar angle $\theta_\gamma^*$ in the CoM frame of the initial $e^+e^-$ system.
At $\mathcal{O} (\alpha_e)$, one has
\cite{Nicrosini:1988hw,Ping:2013jka},
\begin{subequations}
\begin{align}
\mathcal{R}(x_\gamma^*,c_\gamma^*,s_{ee}) 
& \equiv
\Delta \frac{2\alpha_e}{\pi} \frac{1}{1 - c_{\gamma}^{*2}} (x_\gamma^*)^{\kappa -1} \left( 1+ \kappa \ln x_\gamma^* \right)
\nonumber
\\ &
- \frac{\alpha_e}{\pi} \frac{1}{1 -c_\gamma^{*2}} (2-x_\gamma^*),
\\
\Delta 
& \equiv
1 + \frac{\alpha_e}{\pi} \left( \frac{3}{2} \ln \frac{s_{ee}}{m^2_e} + \frac{\pi^2}{3} -2 \right),
\\
\kappa 
& \equiv 
\frac{2\alpha_e}{\pi} \ln \left[ \frac{s_{ee} (1 -c_\gamma^{*2})}{2m^2_e} \right].
\end{align}
\end{subequations}

In our analysis, we evaluate the cross section 
in the small- and large-$x_\gamma$ regions separately
with division by the cutoff parameter $x_{\rm cut}$.
This is because obtaining stable and reliable results
is challenging when performing the integration in the
region $x_\gamma^* \sim 0$ and $|c_\gamma^*| \sim 1$
when using the radiator $\mathcal{R}$.
To address this, in the small-$x_\gamma^*$ region with $x_\gamma^* < x_{\rm cut}$, we employ a reduced radiator, $\mathcal{H}(x_\gamma^*, s_{ee})$, obtained by integrating $\mathcal{R}(x_\gamma^*, c_\gamma^*, s)$ over $c_\gamma^*$.
When $x_{\rm cut}$ is small enough, $\sigma_{\rm ann}^{\rm NLO}$ does not depend on $x_{\rm cut}$.
At $\mathcal{O}(\alpha_e)$, $\mathcal{H}(x_\gamma^*, s_{ee})$ is given by
\cite{Nicrosini:1988hw,Ping:2013jka}
\begin{subequations}
\begin{align}
\mathcal{H}(x_\gamma^*;s_{ee}) 
& \equiv
\Delta  \kappa^\prime (x_\gamma^*)^{(\kappa^\prime -1)} -\frac{1}{2} \kappa^\prime (2-x_\gamma^*),
\\
\kappa^\prime
& \equiv 
  \frac{2 \alpha_e} \pi
\left[ \ln \left(\frac{s_{ee}}{2m^2_e}\right) - 1 \right],
\end{align}
\end{subequations}
and Eq.~(\ref{eq:exc}) reduces to 
\begin{align}
  \frac{d\sigma_{\rm ann}^{\rm NLO}}{d x_\gamma^*}
=
  \mathcal{H}(x_\gamma^*,s_{ee}) \sigma_{\rm ann}^0(s_{\chi\bar{\chi}}).
\end{align}

The total cross section with the ISR effects
for an initial positron of energy $E_{e^+}$
scattering off a stationary atomic electron
in the lab frame is then given by
\begin{align}
& \sigma_{\rm ann}^{\rm NLO} (E_{e^+}) =  \int_0^{x_{\rm cut}} dx_\gamma^* \frac{d\sigma_{\rm ann}^{\rm NLO}}{dx_\gamma^*}
\label{eq:tot-xsec}
\\
& + 
\int^{1-m_e^2/s}_{-1+m_e^2/s} d c_\gamma^* \int_{x_{\rm cut}}^1 dx_\gamma^* \frac{d\sigma_{\rm ann}^{\rm NLO}}{dx_\gamma^* dc_\gamma^*} f_{\rm cut} [E_{\rm miss} (E_{e^+}, x_\gamma^*, c_\gamma^*)],
\nonumber
\end{align}
where $f_{\rm cut}$ denotes the selection criterion
on the missing energy $E_{\rm miss}$ for a given
experiment, arising from the invisible final-state
DM particles.
According to the Lorentz transformation between the 
CoM frame of the $e^+e^-$ system and the lab frame, 
and taking energy conservation into account, 
$E_{\rm miss}$ can be expressed in terms of $x_\gamma^*$, $c_\gamma^*$, and $E_{e^+}$ via
\begin{align}
    E_{\rm miss} = E_{e^+} + m_e - \frac{1}{2}\gamma (1+\beta \, c_\gamma^*)x_\gamma^* \sqrt{s_{ee}},
\end{align}
where $\gamma \equiv (E_{e^+} + m_e)/\sqrt{s_{ee}}$
is the boost factor, with $s_{ee} \equiv 2 m_e (E_{e^+} + m_e)$,
and $\beta \equiv \sqrt{1 - 1/\gamma^2}$.
Note that $x_{\rm cut}$ should be chosen such that the maximum photon energy 
(in the lab frame)
in the first term of \geqn{eq:tot-xsec} 
is much smaller than the energy resolution of the corresponding experiment.
In this work, we adopt $x_{\rm cut} = 0.001$.

\section{\bf Belle II experiment}

In this section, we estimate the contribution with ISR
for the disappearing positron track signature at Belle II.
The disappearing positron track signature arises
from a positron produced from the primary Bhabha scattering,
which subsequently interact with the ECAL to generate invisible 
dark photons  via the bremsstrahlung or annihilation process \cite{Liang:2022pul}.
In the previous study \cite{Liang:2022pul}, only the annihilation process without ISR was considered. 
However, because the energy deposited in the ECAL by the ISR photon is indistinguishable from that caused by the incident positron due to the limited time and position resolution, the ISR contribution must be properly accounted for.

\subsection{\it Signal}

\begin{figure}
\centering
\includegraphics[width=0.48\textwidth]{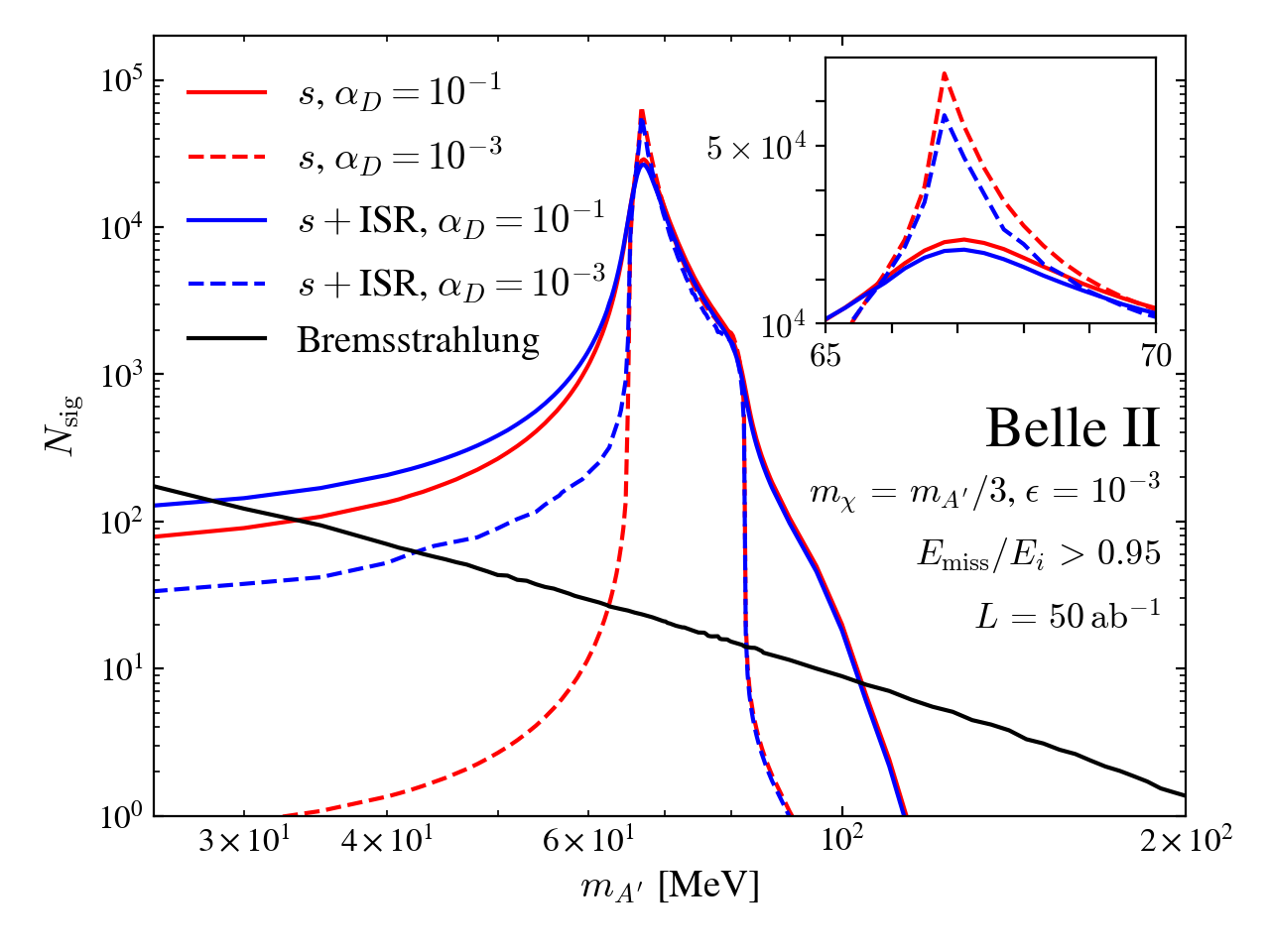} 
\caption{The expected signal event number of the
disappearing positron track signature at Belle II for
the $s$-channel (red),
$s$-channel with ISR
(blue) and
bremsstrahlung process (black), respectively,
in which $m_\chi = m_{A'}/3$, $\epsilon = 10^{-3}$,
and $L = 50 \, {\rm ab}^{-1}$. 
The $\alpha_D$ dependence of each
contribution is illustrated with $\alpha_D = 10^{-1}$
(solid) and $\alpha_D = 10^{-3}$ (dashed).
The inset plot around the resonance peaks shows zoomed-in view of the the dark photon
mass regions of ($65 \sim 70$)\,MeV.
}
\label{fig:aD}
\end{figure}

Taking ISR into account, the event number from the electron-positron annihilation to dark photons 
with energies in the range of 
$(x_{\rm min}, x_{\rm max}) E_i$, where $E_i$ is 
the outgoing positron energy from the Bhabha scattering, 
is
\begin{align}
N_s = 
  n_e \mathcal{L} L_T \int^{E_{\rm max}}_{E_{\rm min}} dE_i \, \frac{d\sigma_B}{dE_i} \sigma_{\rm ann}^{\rm NLO, ME} (E_i),
\label{eq:Ns}
\end{align}
with the selection criterion of 
\begin{align}
f_{\rm cut} = {\rm Boole}[x_{\rm min} < E_{\rm miss}/ E_i < x_{\rm max}].
\end{align}
Here $\mathcal{L} = 50\, {\rm ab}^{-1}$ \cite{Belle-II:2018jsg}
is the expected
integrated luminosity at Belle II and $n_e$ is the
electron number density in the ECAL detector.
The differential cross section $d\sigma_{\rm B}/dE_i$ of the
primary Bhabha scattering, $e^+e^- \rightarrow e^+e^-$,
gives the incident positron energy ($E_i$) spectrum.
Within the barrel coverage of the ECAL, $E_i$ ranges from $E_{\rm min} = 4.35\,\rm GeV$
to $E_{\rm max}=6.62\,\rm GeV$ \cite{Liang:2022pul}. Experimentally,
this incident positron energy $E_i$ can be
reconstructed in the tracker.
For the ECAL CsI crystal in Belle II, the target thickness
$L_T = 16 X_0$ is 16 times of the radiation length
$X_0=1.86 \, {\rm cm}$ \cite{ParticleDataGroup:2018ovx}.
In Ref.\,\cite{Liang:2022pul}, the signal region for the disappearing positron track is defined as a single bin with $(x_{\rm min},x_{\rm max}) = (0.95,1)$.

Note that the annihilation cross section in \geqn{eq:Ns}
is $\sigma_{\rm ann}^{\rm NLO, ME} (E_i)$ rather than
the $\sigma_{\rm ann}^{\rm NLO} (E^+_e)$ in \geqn{eq:tot-xsec}.
At fixed-target configurations, the incident electrons or
positrons can initiate electromagnetic showering in the
target before annihilating with atomic electrons,
resulting in energy loss and the production of secondary
electrons and positrons.
This effect is characterized
by the track-length distribution $T$ \cite{Tsai:1966js}. 
Convoluting $\sigma_{\rm ann}(E_{e^+})$ with $T_e$
then defines an effective cross section that accounts
for the matter effect:
\begin{align}
\hspace{-3mm}
  \sigma_{\rm ann}^{\rm NLO, ME} (E_i)
=
  \int dE_{e^+} \frac {T(E_{e^+}, E_i , L_T)}{L_T}
  \sigma_{\rm ann}^{\rm NLO} (E_{e^+}),
\end{align}
where $L_T$ is the length of the target.
The detailed expressions of the track length
distributions for incident electrons
($T = T_{e}$) and positrons
($T = \overline{T}_{e}$) are
provided respectively in the appendix. 
In addition to NA64 and Belle II, our formalism can also
apply to the resonant dark photon searches at future
electron- or antimuon-on-target experiments such as
LDMX \cite{LDMX:2018cma}, DarkSHINE \cite{DarkSHINE:2022mak},
and DREAMuS \cite{Xu:2025spd,DREAMuS2025}.

\begin{figure}[t]
\centering
\includegraphics[width=0.48\textwidth]{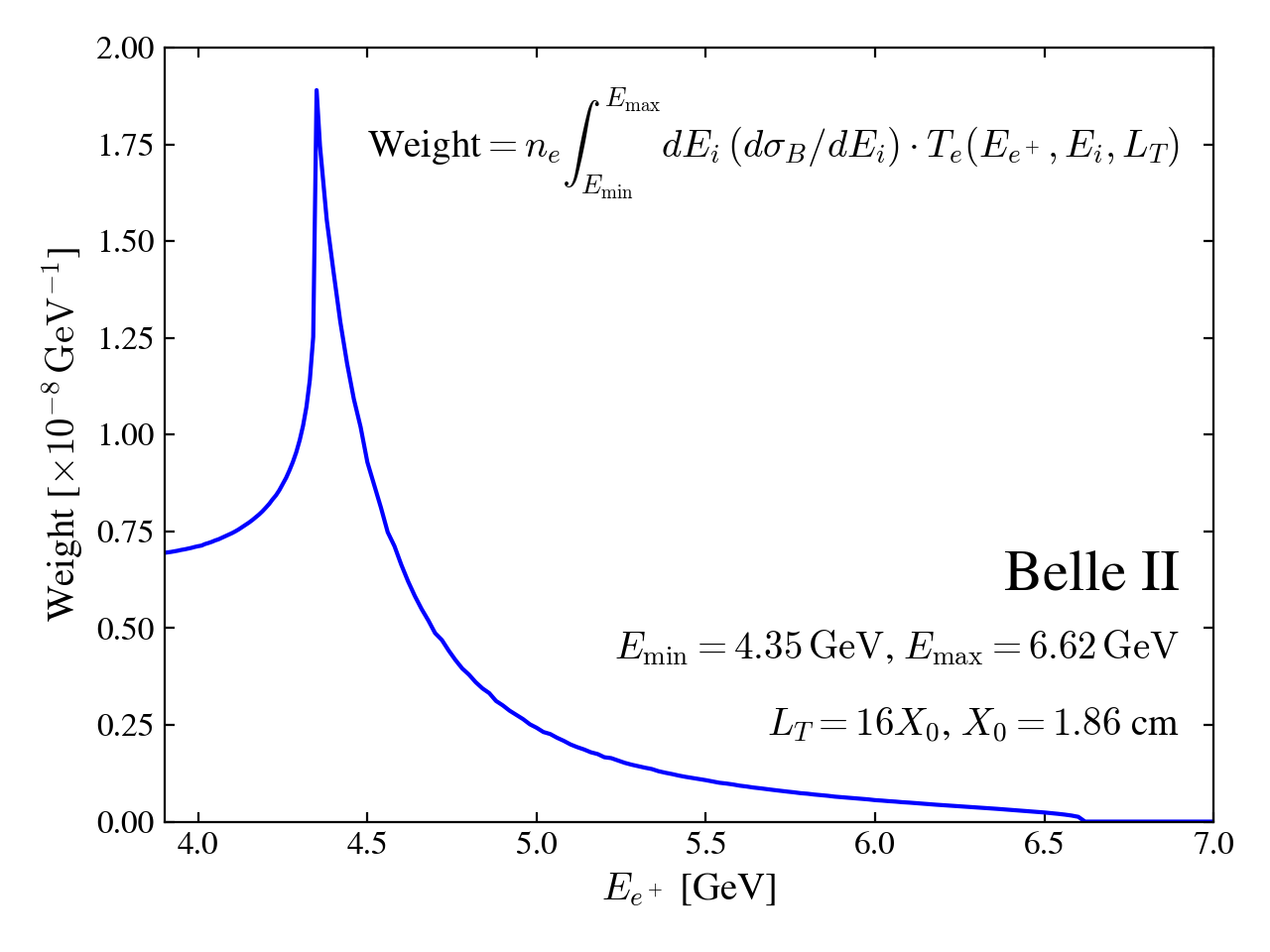} 
\caption{Weighting factor for the signal event number
given by convoluting the positron track-length
distribution $T_e(E_{e^+}, E_i, L_T)$ and the differential
Bhabha scattering cross section $d \sigma_{\rm B} / d E_i$
over the outgoing positron energy range,
$4.35\,{\rm GeV} \leq E_i \leq 6.62\,{\rm GeV}$,
from the Bhabha scattering at Belle II.
}
\label{fig:weight}
\end{figure}

The expected signal event numbers for the annihilation
processes with and without ISR, as well as for the
bremsstrahlung process, within the signal region are
shown in \gfig{fig:aD}, assuming $m_{\chi} = m_{A'}/3$,
$\epsilon = 10^{-3}$, and an integrated luminosity
of $\mathcal{L} = 50~\mathrm{ab^{-1}}$ for illustration.
Since the annihilation processes, both with and without ISR, 
provide resonant production for the dark photon $A'$,
their event spectra exhibit a peak.
The resonant region
${65\,\rm MeV} \lesssim \sqrt{s} \lesssim 82\,\rm MeV$
is fully consistent with $s_{ee} = 2 m_e (E_{e^+} + m_e)$
where the initial positron energy is in the range
of $0.95 \times 4.35\,\mbox{GeV} \leq E_{e^+} \leq 6.62$\,GeV
as $E_{e^+}$ larger than $95\%$ of $E_i$ is considered.
Especially, the peak appears around 
$m_{A'} \approx 67$\,MeV.
This is because the incoming
positron spectrum
$\int T_e(E_{e^+}, E_i, L_T) d E_{e^+} (d \sigma_{\rm B}/d E_i)$,
as shown in \gfig{fig:weight}, peaks at
$E_{e^+}=E_{\rm min} = 4.35$\,GeV due to convolution
of the increasing track-length distribution
$T_e$ and the decreasing Bhabha scattering cross section $d \sigma_{\rm B} / d E_i$.

On the other hand, the bremsstrahlung event rate
decreases with the dark photon mass.
Although there is
also a resonance around 
$E_{e^+} \approx m_{A'}$, as implied by
Eq.(4.6) of \cite{Liang:2022pul},
it lies outside the region of 
interest for the present analysis.

\begin{figure}[t]
\centering
\includegraphics[width=0.48\textwidth]{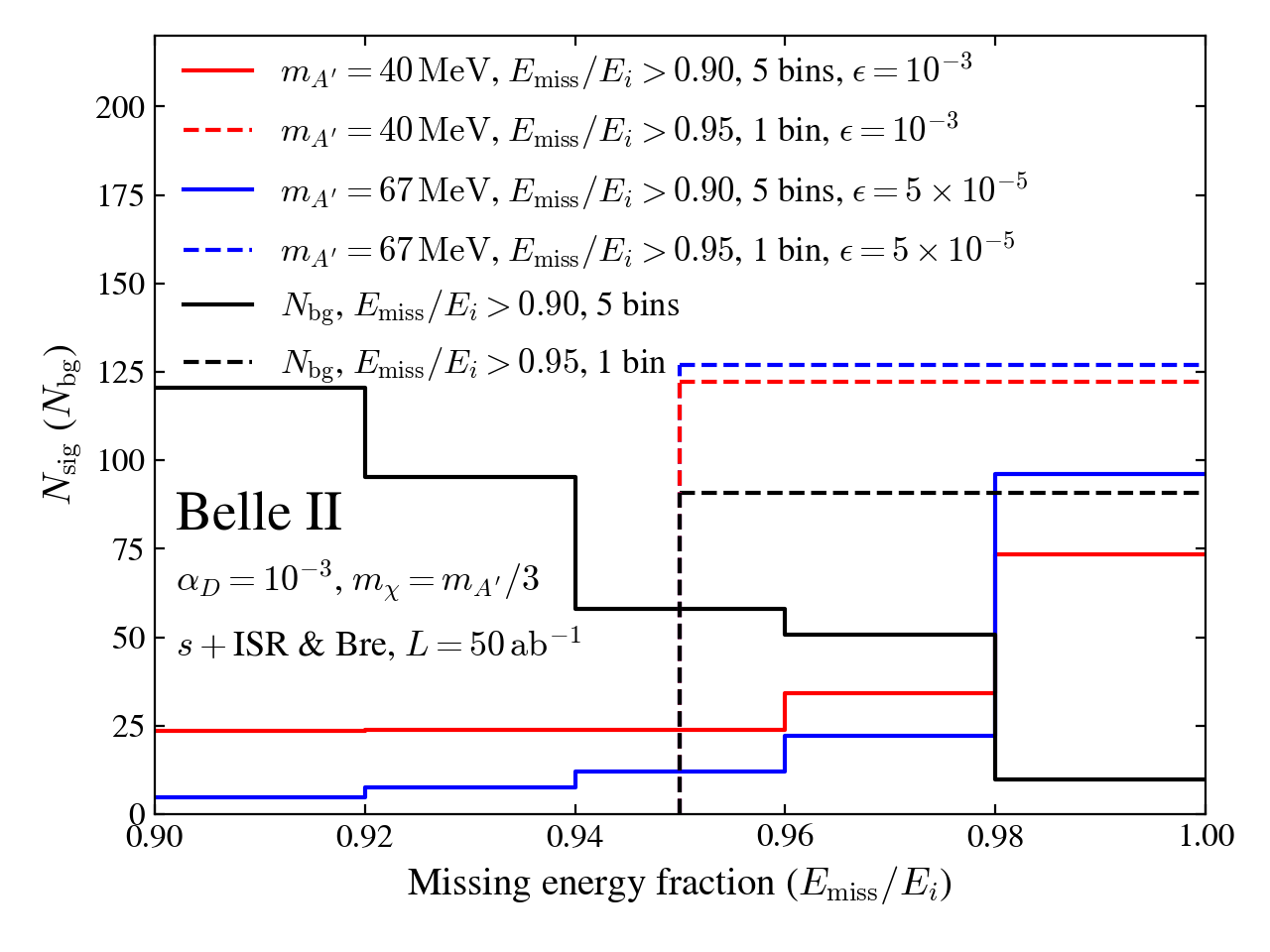}
\caption{The expected signal (colorful lines) and
background (black lines) event rates at Belle II with an
integrated luminosity $L = 50 \, {\rm ab}^{-1}$.
The analysis takes two fixed dark photon masses,
$m_{A'} = 40 \, {\rm MeV}$ (red lines) and
$m_{A'} = 67 \, {\rm MeV}$ (blue lines).
To show the signal and background with similar size,
the signal event rates have been scaled by tuning
the dark photon kinetic mixing parameter
$\epsilon = 10^{-3}$ for $m_{A'} = 40$\,MeV
and
$\epsilon = 5\times 10^{-5}$
for $m_{A'} = 67$\,MeV.
The solid lines denote the event numbers estimated
for the binning with a size of $2\%$ for the
missing energy fraction $x = E_{\rm miss}/E_i$,
ranging from $90\%$ to $100\%$. For the dashed
lines, however, the event numbers are obtained
within a single bin covering the missing energy
fraction from $95\%$ to $100\%$.
}
\label{fig:eventRate}
\end{figure}

Note that the $s$-channel spectrum shape
depends on $\alpha_D$.
As pointed out in the previous section, the dark
photon decay width can enter the $s$-channel propagator
with a positive virtuality ($s > 0$).
Then the $s$-channel resonance would be subject to
the decay width
$\Gamma_{A' \rightarrow \chi \chi}$ from \geqn{eq:xsec-s-chan},
which 
scales linearly with the dark fine-structure constant
$\alpha_D$. 
For illustration, we present the
corresponding results for $\alpha_D = 10^{-1}$, 
and $10^{-3}$, respectively, in \gfig{fig:aD}.
A larger $\alpha_D$ corresponds to a wider decay width of
the dark photon and allows the $s$-channel obtained at leading order to probe a broader mass region. 
For the extremely small coupling
$\alpha_D \lesssim 10^{-3}$ that leads to a
negligibly small decay width, the dark photon mass
region that the $s$-channel can probe coincides with
the CoM energy determined by the energy
range of the incident positrons, namely
${65\,\rm MeV} \lesssim \sqrt{s} \lesssim 82\,\rm MeV$.
As shown in \gfig{fig:aD}, for a sufficiently small dark gauge coupling ($\alpha_D = 10^{-3}$), the ISR correction significantly distorts the resonance shape with only $s$-channel contribution: the peak becomes flatter with fewer events in the narrow resonance region, while the number of events in the low-mass region (${40\,\rm MeV} \lesssim m_{A'} \lesssim 65\,\rm MeV$) increases by up to two orders of magnitude.
On the other hand, the ISR effect
becomes relatively less important for a large enough $\alpha_D = 0.1$
by comparing the red and blue solid curves in \gfig{fig:aD}.

\subsection{\it Background}

Neutral particles, including photons and neutrons,
are abundantly
produced when positrons strike the ECAL. 
For the disappearing positron track signature at
Belle II, the SM backgrounds arise when a large
fraction of the positron energy is carried away by
these neutral particles, which subsequently penetrate
the $K_L$-muon detector (KLM) located behind the ECAL.
We estimate the neutron-induced backgrounds at
Belle~II using a \textsc{Geant4}
\cite{GEANT4:2002zbu} simulation, following the
procedure described in \cite{Liang:2022pul}.
In our analysis, we consider events that have
large missing energy in the range of 
$x \equiv E_{\rm miss}/E_i>90\%$.
We further bin those events with a bin width of 2\%.

\begin{table}[t]
\centering
\begin{tabular}{c|cc|c}
$E_{\rm miss}/E_i$ ($\%$)
& $N^n_{\rm bg}$ & $N^\gamma_{\rm bg}$ & $N_{\rm bg}$  \\
\hline
$90 \sim 92$ & 111 & 9.6 & 120.6 \\
$92 \sim 94$ & 87  & 8.3 & 95.3 \\
$94 \sim 96$ & 51  & 7.0 & 58.0 \\
$96 \sim 98$ & 45  & 5.6 & 50.6 \\
$98 \sim 100$ & 6  & 3.9 & 9.9 \\
\hline
$95 \sim 100$ & 78 & 12.8 & 90.8 \\
\end{tabular}
\caption{The background event numbers induced by
photon and neutron in each $2\%$ missing energy
fraction bin 
$x \equiv E_{\rm miss}/E_i$
at Belle II.
The total number of backgrounds is obtained as
$N_{\rm bg} \equiv N^n_{\rm bg} + N^\gamma_{\rm bg}$. 
}
\label{tab:BG_Belle2}
\end{table}

The photon-induced backgrounds can be analytically estimated without complicated simulation.
The energy distribution of photons produced in the electromagnetic shower initiated by an incident positron in the ECAL
is \cite{Tsai:1966js}
\begin{align}
     \frac{dN_\gamma}{d x_\gamma} (t,x_\gamma) &= \frac{1}{x_\gamma} \frac{(1-x_\gamma)^{\frac{4}{3}t}- e^{-\frac{7}{9}t}}{\frac{7}{9} + \frac{4}{3} \ln (1-x_\gamma)}~, 
\end{align}
where $x_\gamma \equiv E_\gamma/E_i$
denotes the energy fraction carried by the photon.
Thus, the number of photons exiting the ECAL boundary within each energy bin
 can be estimated as
\begin{align}
  N^\gamma_{\rm bg}
\equiv
  \epsilon^\gamma_{\rm KLM} N_{e^+}
  \int_{x_{\rm min}^i}^{x_{\rm max}^i}
  \frac{dN_\gamma}{d x_\gamma} (t = 16,x_\gamma)
  dx_\gamma,
\end{align}
where $x^i_{\rm min}$ ($x^i_{\rm max}$) is the minimum (maximum) energy fraction in the $i$-th bin,
and $N_{e^+} = 6\times 10^{11}$ is the total number of positrons entering the ECAL.
The KLM veto efficiency for photons, $\epsilon^\gamma_{\rm KLM}$, is estimated using the IFR veto efficiency at BABAR to be approximately $4.5 \times 10^{-4}$~\cite{BaBar:2008aby} as a conservative estimation.

The event numbers of neutron- and photon-induced backgrounds, as well as the total background in each 2\% energy fraction bin, 
are given in \gtab{tab:BG_Belle2}.  
The last row corresponds to the number of background
events collected within the range $x > 95\%$
that was considered as the signal region in the previous
study \cite{Liang:2022pul} for comparison.  

\subsection{\it Sensitivity}

The energy measurement at Belle II can be as good
as $\delta E/E \simeq 2\%$ \cite{Belle-II:2018jsg}.
Although the single-bin analysis in
\cite{Liang:2022pul} 
already achieves strong background suppression, 
it does not exhaust
the information. As shown in \gfig{fig:eventRate}
and \gtab{tab:BG_Belle2},
the background event rate decreases fast with
the increasing fraction of invisible energy.
For comparison, the signal event rate increases
with the missing energy fraction $x$ as illustrated
with red and blue lines in \gfig{fig:eventRate}. 
With a smaller bin size, one would expect
the ratio of signal to background events to increase.

\begin{figure}[t]
\centering
\includegraphics[width=0.48\textwidth]{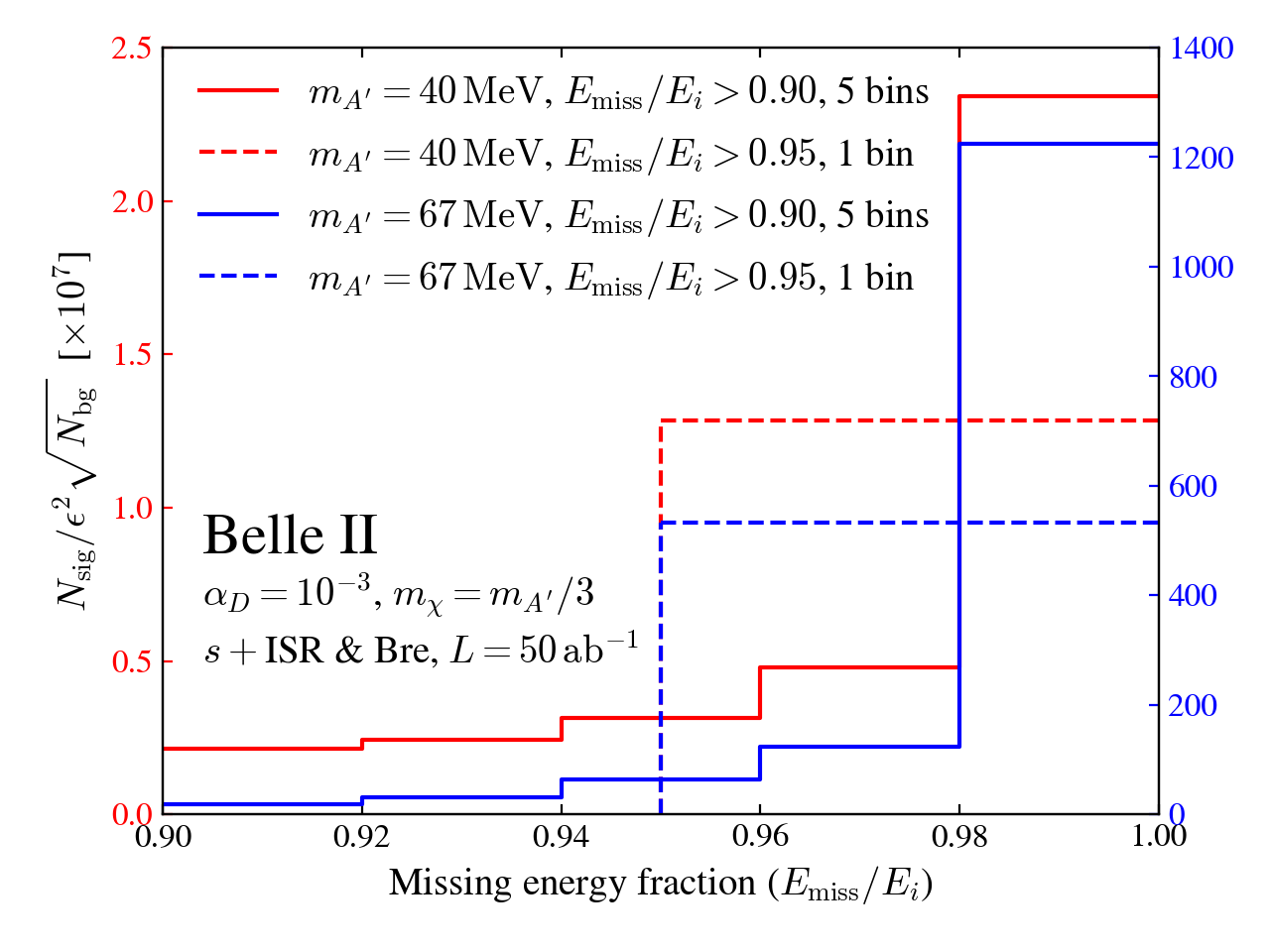} 
\caption{The expected sensitivity contribution as
a function of missing energy fraction,
normalized by $\epsilon^2$ at Belle II with an
integrated luminosity $L = 50 \, {\rm ab}^{-1}$.
The analysis takes two fixed dark photon masses,
$m_{A'} = 40 \, {\rm MeV}$ (red lines) and
$m_{A'} = 67 \, {\rm MeV}$ (blue lines).
For the solid lines, the sensitivity is estimated
for each bin with a bin size of $2\%$ for the
missing energy fraction 
$x = E_{\rm miss}/E_i$,
ranging from $90\%$ to $100\%$. For the dashed
lines, however, the sensitivity contribution is obtained
within a single bin covering the missing energy
fraction from $95\%$ to $100\%$.
}
\label{fig:Belle_binned_analysis}
\end{figure}

In addition, the major difference between 
the $s$-channel annihilation processes without and with ISR
is that in the former
case the dark photon takes away all the positron
energy while in the latter case it 
can carry only a part of it.
So the multiple-bin analysis 
not only helps
to extract the spectrum information but also  
enables a discrimination between the annihilation
processes that occur with and without ISR.
One may expect the
multi-bin analysis to significantly enhance the
sensitivity.

With the assumption that the dominant uncertainty of the missing energy searches at the ECAL detector mainly comes from backgrounds, the expression of $\chi^2$ is approximately given by summation of the squared signal ratios,
\begin{align}
    \chi^2 \simeq \sum_{i} \left( \frac{N^{(i)}_{\rm sig}}{\sqrt{N^{(i)}_{\rm bg}} }\right)^2~,
\label{eq:chi_equare_Belle_II}
\end{align}
where $N^{(i)}_{\rm sig}$ ($N^{(i)}_{\rm bg}$) is the
number of signal (background) events in the $i$-th
bin of the missing energy fraction. To obtain the $90\%$ CL limits on the kinetic mixing parameter, the required value of $\chi$ is calculated from
\begin{align}
    \int^{\infty}_\chi d\chi' \, \sqrt{\frac{2}{\pi}}e^{-\chi'^2/2} = 0.1,
\end{align}
which gives $\chi = 1.645$. Then sensitivity bounds on the kinetic mixing parameter are obtained by solving the equation $\chi(\epsilon) = 1.645$ as $\chi^2$ is parametrized by the single parameter $\epsilon$. 

\begin{figure}
\centering
\includegraphics[width=0.48\textwidth]{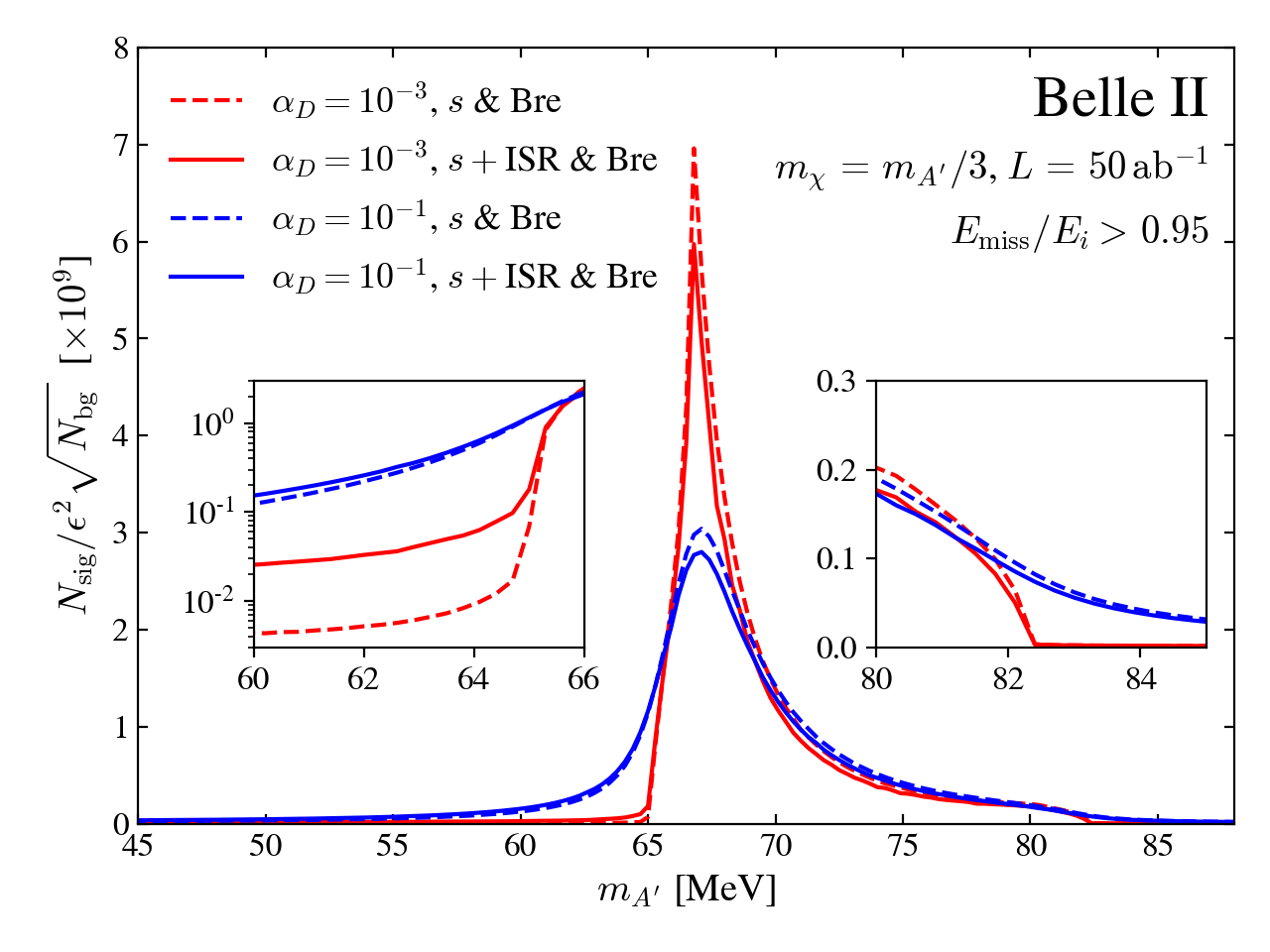} 
\caption{The expected sensitivity contribution at Belle II as a function of the dark photon mass $m_{A'}$, normalized by $\epsilon^2$ for $\alpha_D = 10^{-3}$ (red) and $\alpha_D = 10^{-1}$ (blue) with $m_\chi = m_{A'}/3$ and an integrated luminosity of $L = 50 \, {\rm ab}^{-1}$.
While the dashed curves account for contributions
from $s$-channel and bremsstrahlung processes,
the solid curves consider the $s$-channel contributions
with ISR by convoluting the electron partonic
distribution function as well as bremsstrahlung process contributions.
Two inset plots show zoomed-in views of the
dark photon mass regions of $(60 \sim 66) \, {\rm MeV}$
and $(80 \sim 85) \, {\rm MeV}$, respectively.}
\label{fig:sensitivity_Bell_varying_aD}
\end{figure}

\gfig{fig:Belle_binned_analysis} shows the
sensitivity contribution, estimated with
$N_{\rm sig} / \epsilon^2 \sqrt{N_{\rm bg}}$
from each bin. Since the signal event numbers
universally scale with the kinetic mixing
parameter $\epsilon^2$, we choose to remove
it for easy comparison. Since the background
event rate decreases with the missing energy
fraction $x$, one would expect the major
contribution to come from the final bin
$0.98 \leq x \leq 1$. The single-bin analysis
is also shown with dashed lines for comparison.
One may see that the multi-bin analysis can
truly significantly enhance the sensitivity
as expected. The single final bin with
$0.98 \leq x \leq 1$ is already much better
than the combined $0.95 \leq x \leq 1$ since
the background event number decreases by almost
a factor of 10 as shown in \gtab{tab:BG_Belle2}.

\gfig{fig:sensitivity_Bell_varying_aD} illustrates
the significant improvement in the signal
sensitivity with $m_\chi = m_{A'}/3$. The dashed curves are obtained with only the $s$-channel
and bremsstrahlung contributions while the solid ones
account for the $s$-channel contribution with
ISR as well as the bremsstrahlung counterpart.
One can see that the enhancement can even reach
approximately a factor of ten for a smaller dark
coupling $\alpha_D = 10^{-3}$ (red solid curve) which is
especially true on the left side of the sensitivity
peak as made explicit in the left inset plot.
This is because the ISR contribution can
dominate over its $s$-channel counterpart if the
dark photon decay width is small as already shown
in \gfig{fig:aD}. For comparison, the improvement
on the right side of the sensitivity peak is quite
minor since the ISR contribution there
never dominates even for the tiny dark coupling
$\alpha_D = 10^{-3}$.

\begin{figure}[t]
\centering
\includegraphics[width=0.48\textwidth]{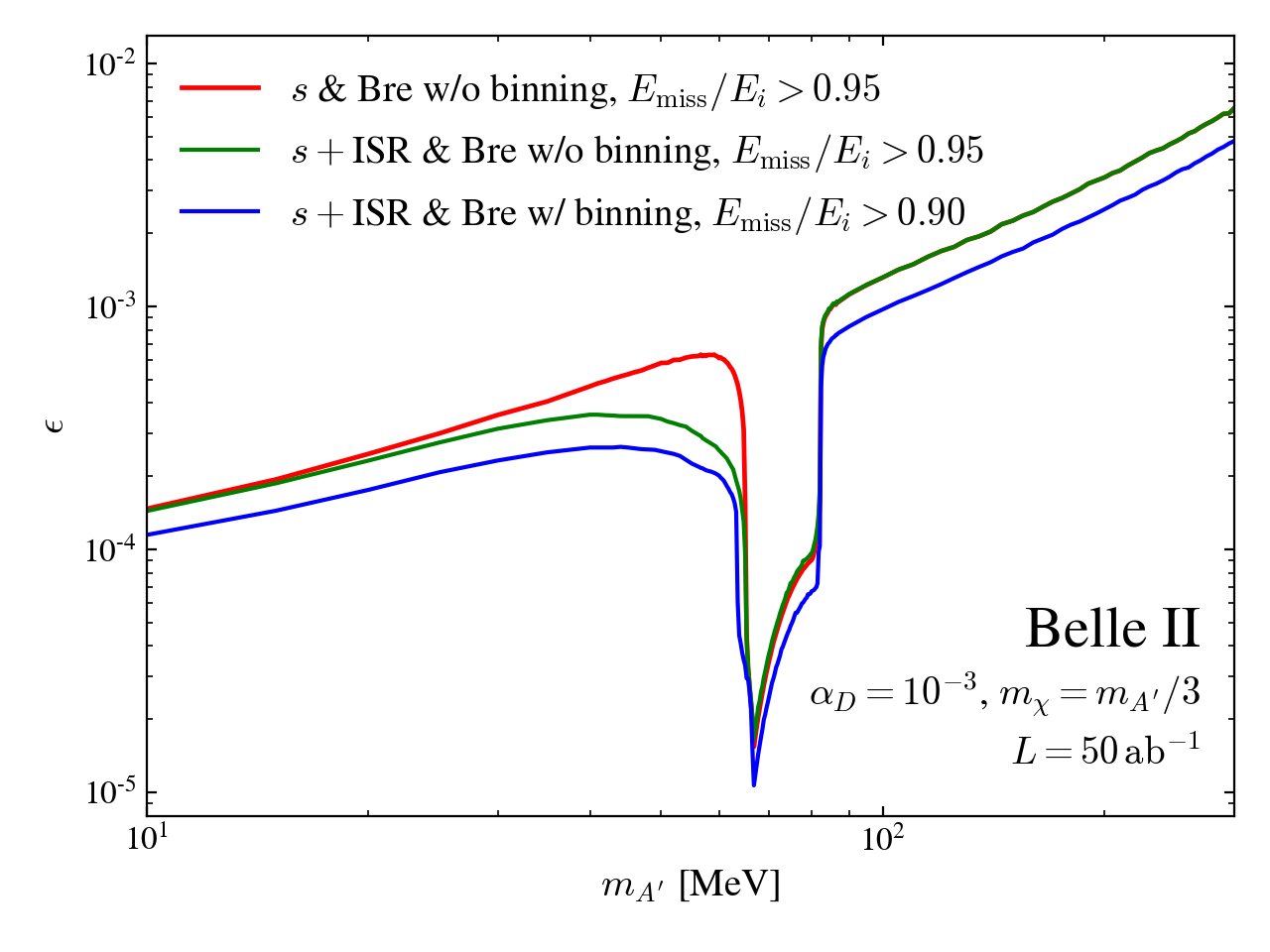} 
\caption{Predicted sensitivities to the kinetic mixing parameter $\epsilon$ as a function of the dark photon mass $m_{A'}$ at Belle II. The red curve represents the sensitivity from the $s$-channel and bremsstrahlung processes for the dark photon production without binning.
The green curve
accounts for the $s$-channel contribution with ISR by convoluting the electron partonic distribution function as well as the bremsstrahlung process contribution. The blue curve corresponds to a binned analysis incorporating the $s$-channel with ISR
and bremsstrahlung processes in the energy fraction range from $90\%$ to $100\%$ with a bin size of $2\%$. 
}
\label{fig:sensitivity_Belle}
\end{figure}

\gfig{fig:sensitivity_Belle} presents the sensitivity bounds on the kinetic mixing parameter with and without binned analysis,
in which $\alpha_D = 10^{-3}$ and $m_{A'} = 3 m_\chi$.
The red curve, which accounts only for contributions
from the $s$-channel at leading order
and bremsstrahlung processes in
the dark photon production, is consistent with the
result from \cite{Liang:2022pul}. 
The sensitivity curve has a dip in the dark photon mass
region from 65\,MeV to 82\,MeV arising
from the resonant production of a dark photon via the
$s$--channel process with
$m^2_{A'} = s_{ee} = 2m_e(E_{e^+}+m_e)$.
This dip
is consistent with the signal spectrum in \gfig{fig:aD}
for $\alpha_D = 10^{-3}$. Further
convoluting the annihilation process with the electron parton distribution function
would significantly
enhance the sensitivity for the dark photon mass
below 65\,MeV at Belle II from the red
curve to the green one in \gfig{fig:sensitivity_Belle}.

The sensitivity to the kinetic mixing parameter
$\epsilon$ can be further enhanced by binning
the missing energy fraction. 
As shown in \gfig{fig:eventRate}, 
the signal increases with the missing energy fraction, 
whereas the background exhibits a decreasing trend.
The multi-bin analysis can fully
employ these opposite 
trends to enhance the
sensitivity contribution, as shown in 
\gfig{fig:Belle_binned_analysis}. This is true
even when only the $s$-channel
at leading order and bremsstrahlung
contributions are considered. 
Since the $s$-channel
leads to almost full missing energy, the multi-bin
analysis can effectively separate the $s$-channel
contribution from its bremsstrahlung counterpart
to some extent. 
This explains why the multi-bin improvement
appears in almost the entire
mass range. From the green curve to the blue one in \gfig{fig:sensitivity_Belle}, 
both including the ISR effect,
the sensitivity is globally enhanced
by a factor of 1.3.
In addition, the resonant region is also slightly
extended to ${63\,\rm MeV} \lesssim m_{A'} \lesssim 82\,\rm MeV$
because the energy range of the incident positron is
${0.90 \times 4.35\,\rm GeV} < E_{e^+} < 6.62 \,\rm GeV$
with $E_{e^+}$ being larger than $90\%$ of $E_i$ is considered in
the multi-bin analysis.

The mono-photon search for $e^+e^- \rightarrow \gamma A'$ at the Belle II collider also provides a sensitivity bound on the kinetic mixing parameter~\cite{Belle-II:2018jsg}. The expected sensitivity curve shown in FIG. 209 of~\cite{Belle-II:2018jsg} could be improved by increasing the integrated luminosity from $20 \, {\rm fb}^{-1}$ to $50 \, {\rm ab}^{-1}$. However, the characterization of the cosmic ray background, which significantly impacts the sub-GeV region of the dark photon mass, remains incomplete. Therefore, the projected sensitivity below about $2\, {\rm GeV}$ is not presented in FIG. 11 of~\cite{Belle-II:2022cgf}. Given the effect of the cosmic ray background on the mono-photon search in the sub-GeV region, we do not explicitly include its corresponding bound in our results.

\section{NA64 experiment}

For Belle II, the injecting positron $e^+$ 
effectively acts as a dumped beam in
the detector for the dark photon
search. It would be of great advantage to have
a specifically designed fixed-target experiment
such as NA64 \cite{Andreas:2013lya}.
Note that the NA64 experiment operates in
three modes to search for dark photons
\cite{Banerjee:2019pds, Andreev:2021fzd, NA64:2023ehh},
\begin{itemize}
\item (a) 100\,GeV incident electrons with $N_{\rm EoT} = 2.8 \times 10^{11}$;
\item (b) 100\,GeV incident positrons with $N_{\rm EoT} = 1.0 \times 10^{10}$;
\item (c) 60\,GeV and 40\,GeV incident positrons, each with $N_{\rm EoT} = 1.0 \times 10^{11}$,
\end{itemize}
where $N_{\rm EoT}$ is the number of electrons or positrons on target.
The first two modes have already collected data, while the third mode is planned for future runs \cite{NA64:2023ehh}.

Taking ISR into account, the number of signal events arising from the annihilation process at NA64 is given by
\begin{align}
&
N_{s}
=
\epsilon_d n_e L_T N_{\rm EoT} \sigma_{\rm ann}^{\rm NLO, ME} (E_i), 
\end{align}
where $\epsilon_d \simeq 0.5$ denotes the detector efficiency,
In addition, the target thickness, $L_T = 40 X_0$ is
40 times of the radiation length $X_0 = 0.56\ \text{cm}$
for the NA64 ECAL which is primarily composed of lead.

The signal region adopted by the NA64 collaboration
for the missing energy search is defined as
$E_{\rm ECAL} < 0.5 E_i$, within which no events
were observed  \cite{Banerjee:2019pds},
where $E_{\rm ECAL}$ denotes the
energy deposited in the ECAL.
This exceptionally clean signal region is enabled by
the delicate design of the NA64 experiment that
features an ECAL with a depth of approximately
40 radiation lengths as the target, a massive and hermetic hadronic calorimeter with about 30 nuclear interaction
lengths, and a large high-efficiency veto counter
placed between the two calorimeters
\cite{Banerjee:2019pds}.


According to the Poisson distribution, the probability
of observing $k$ events with the expectation value
$\lambda$ is
\begin{align}
  P(k; \lambda)
=
  \frac{\lambda^k e^{-\lambda}}{k!}.
\end{align}
With the null observation ($n_{\rm obs}=0$) and the vanishing background ($\lambda=N_{\rm sig}$),
the 90\% confidence level constraint on the signal event number $N_{\rm sig}$ is
\begin{align}
   \sum^{n_{\rm obs}}_{k=0} P(k,N_{\rm sig}) = 0.1,
\end{align}
which gives $N_{\rm sig} = 2.30$.

\begin{figure}[t]
\centering
\includegraphics[width=0.48\textwidth]{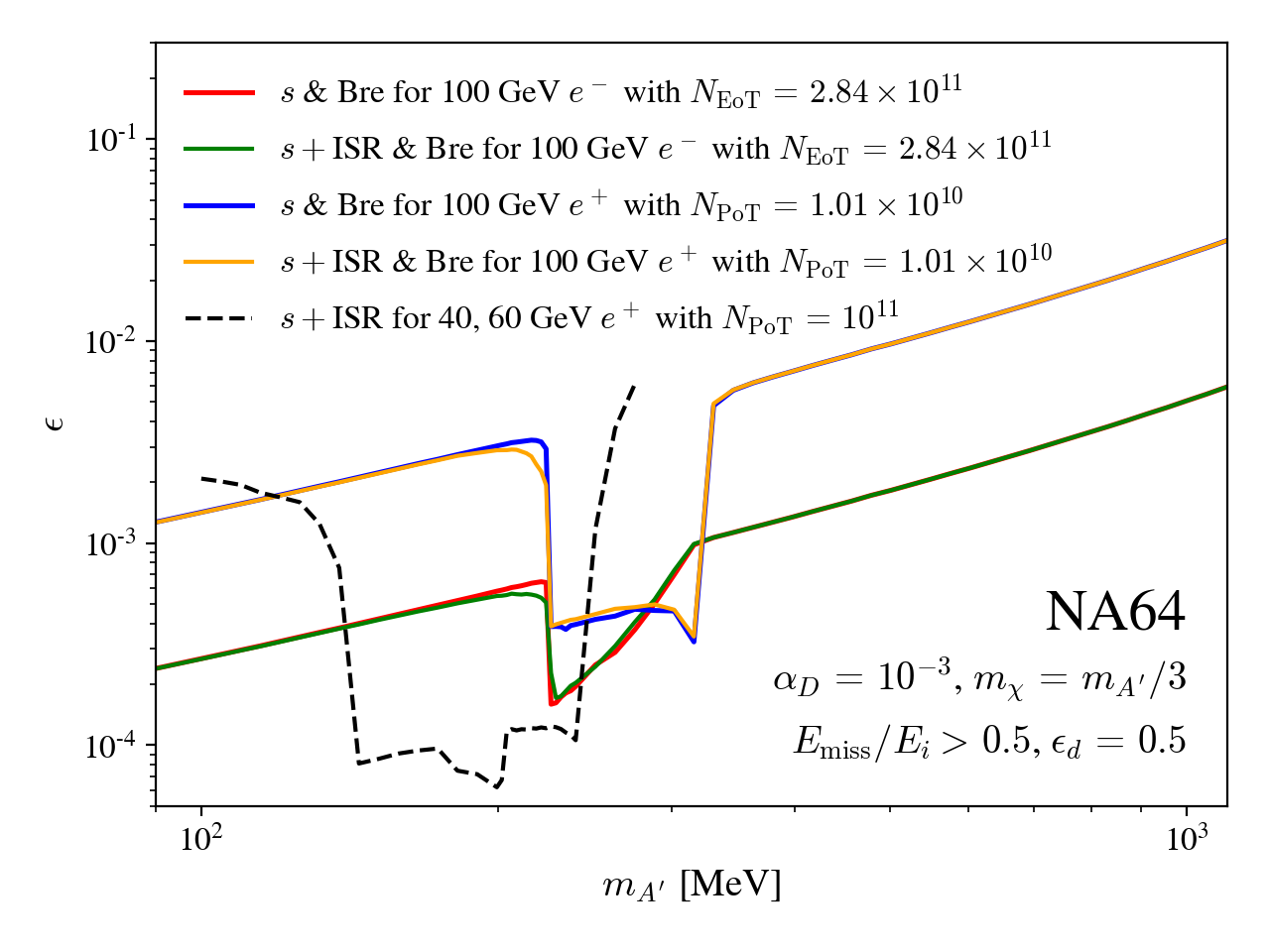} 
\caption{Predicted sensitivities to the kinetic mixing parameter $\epsilon$ as a function of the dark photon mass $m_{A'}$ at NA64. The signal events are selected in the energy fraction range 
$E_{\rm miss}/E_i$
from $50\%$ to $100\%$. The red curve shows the sensitivity
for the 100\,GeV electron beam
including contributions from the $s$-channel and bremsstrahlung processes, 
while the green curve accounts for the $s$-channel contribution with ISR as well as the bremsstrahlung process contribution, leading to sensitivity enhancement.
The sensitivities obtained from the 100\,GeV positron beam with and without ISR effects are shown as the orange and blue curves, respectively. The sensitivity for a future
measurement at two different positron beam energies, 60 and 40 GeV, within the $100-300$\,MeV dark photon mass window is shown as the black dashed curve.
}
\label{fig:sensitivity_NA64}
\end{figure}

\gfig{fig:sensitivity_NA64} presents the resulting
sensitivity curves on the kinetic mixing parameter at NA64
for the dark photon mass from 100\,MeV to
1\,GeV. The red curve is obtained from the
$s$-channel at leading order
and bremsstrahlung contributions only,
whose peak near 200\,MeV arises from the resonant
dark photon production in the $s$-channel.
As discussed in the previous sections, the resonant
dark photon production in $s$-channel appears at
$m^2_{A'} = s_{ee} = 2 m_e (E_{e^+}+m_e)$.
Since the track-length distribution $\overline{T}_e$ 
is a decreasing function of the positron energy 
$E_{e^+}$ in the energy range from 
$0.5 E_i$ to $E_i$, the peak
on the sensitivity curve is located at
$m_{A'} = \sqrt{2 m_e (0.5 E_i)} = 226 \,{\rm MeV}$
and the dark photon can be resonantly produced until
its mass reaches 
$m_{A'} = \sqrt{2 m_e E_i} = 320 \,{\rm MeV}$.

For comparison, the green curve
accounts for the ISR effect to the annihilation processes.
Since the extra photon
in the final state takes away some energy, the dark
photon with mass below 226\,MeV can also be produced
on shell in the presence of
the initial-state radiation.
This leads to enhancement of the sensitivity in the
mass region below the peak position compared to the contribution of the $s$-channel at leading order.

The orange and blue curves represent the sensitivities obtained using the positron beams, with and without accounting for the ISR effect. Similar to the case of the electron beam experiments, the dark photon can be resonantly generated within the mass region of $226-320 \, {\rm MeV}$. The relatively flat behavior of the sensitivity curves arises from the track-length distribution of an incident positron. For the same reason, the sensitivity enhancement appears below the peak region due to the ISR effect when comparing between blue and orange curves.

The sensitivity using low energy positron beams with the larger number of positrons on target at future program is denoted as the black dashed curve. It accounts for the incident positrons with their beam energy of $40 \, {\rm GeV}$ and $60\,{\rm GeV}$ at $N_{\rm PoT}=10^{11}$ for each energy point. The peak region appears from $m_{A'} = \sqrt{2 m_e (0.5 \times 40\, {\rm GeV})} = 143 \,{\rm MeV}$ to $m_{A'} = \sqrt{2 m_e (60\, {\rm GeV})} = 248 \,{\rm MeV}$ by combining the number of signal events from two energy points.
With combination of two beam energies, the
sensitivity curve has a dip in the middle of the bottom now.

\vspace{2mm}

\section{\bf Conclusions}

In this work, we investigate the contribution of
the initial-state radiation $e^+ e^- \rightarrow \gamma A'^{(*)} $
to the invisible dark photon searches.
We demonstrate that the sensitivity to the kinetic
mixing parameter below the resonant region of the
dark photon mass can be enhanced because the additional
photon of the initial state radiation carries away energy.
For illustration,
we consider both the disappearing positron track signature
at Belle II and the large missing energy search at NA64.
Due to the radiative return feature that effectively reduces the center-of-mass energy, 
the $s$-channel contribution with ISR
dominates over its $s$-channel without ISR and bremsstrahlung counterparts
when the dark photon mass
is slightly below the resonance peak. The resulting enhancement
is
significant for a small dark fine-structure constant
$\alpha_D = g_D^2/(4\pi) = 10^{-3}$
with $g_D$
denoting the dark photon–fermion coupling.
For the Belle II case, using the same signal region
as \cite{Liang:2022pul}, we find that
the ISR effect to the annihilation process
can enhance the sensitivity to the kinetic
mixing parameter $\epsilon$
by a factor of 1.3 to 2.7
for the dark photon mass range
${40\,\rm{MeV}} \lesssim m_{A'} \lesssim {60\,\rm{MeV}} $.

Furthermore, a binned analysis incorporating the
energy resolution of the Belle II calorimeter can
improve its sensitivity
by 30\%
across the
entire mass range. 
On one hand, binning the missing
energy fraction can fully employ the decreasing background
and the increasing signal event spectra to enhance
the contrast. On the other hand, the multi-bin analysis
can effectively separate 
the annihilation processes with resonant behaviors
from the non-resonant bremsstrahlung
process.
For the NA64 case,
the ISR contribution to the annihilation process
can still yield a 
noticeable improvement in sensitivity
for a dark photon mass around 200 MeV, even though
the positrons are
secondarily produced in the electromagnetic shower
initiated by the incident electron.

\vspace{2mm}

\section*{Acknowledgements}
SFG and UM are supported by the National Natural Science
Foundation of China (12375101, 12425506, 12090060 and 12090064) and the SJTU Double First
Class start-up fund (WF220442604).
ZL is supported by the National Natural Science Foundation of China 12275128.
JL is supported by the National Natural Science
Foundation of China 12347121.
SFG is also an affiliate member of Kavli IPMU, University of Tokyo.

\vspace{2mm}

\section{\bf Appendix: Track-Length Distribution}
\label{sec:app}

\begin{figure}[t]
\centering
\includegraphics[width=0.48\textwidth]{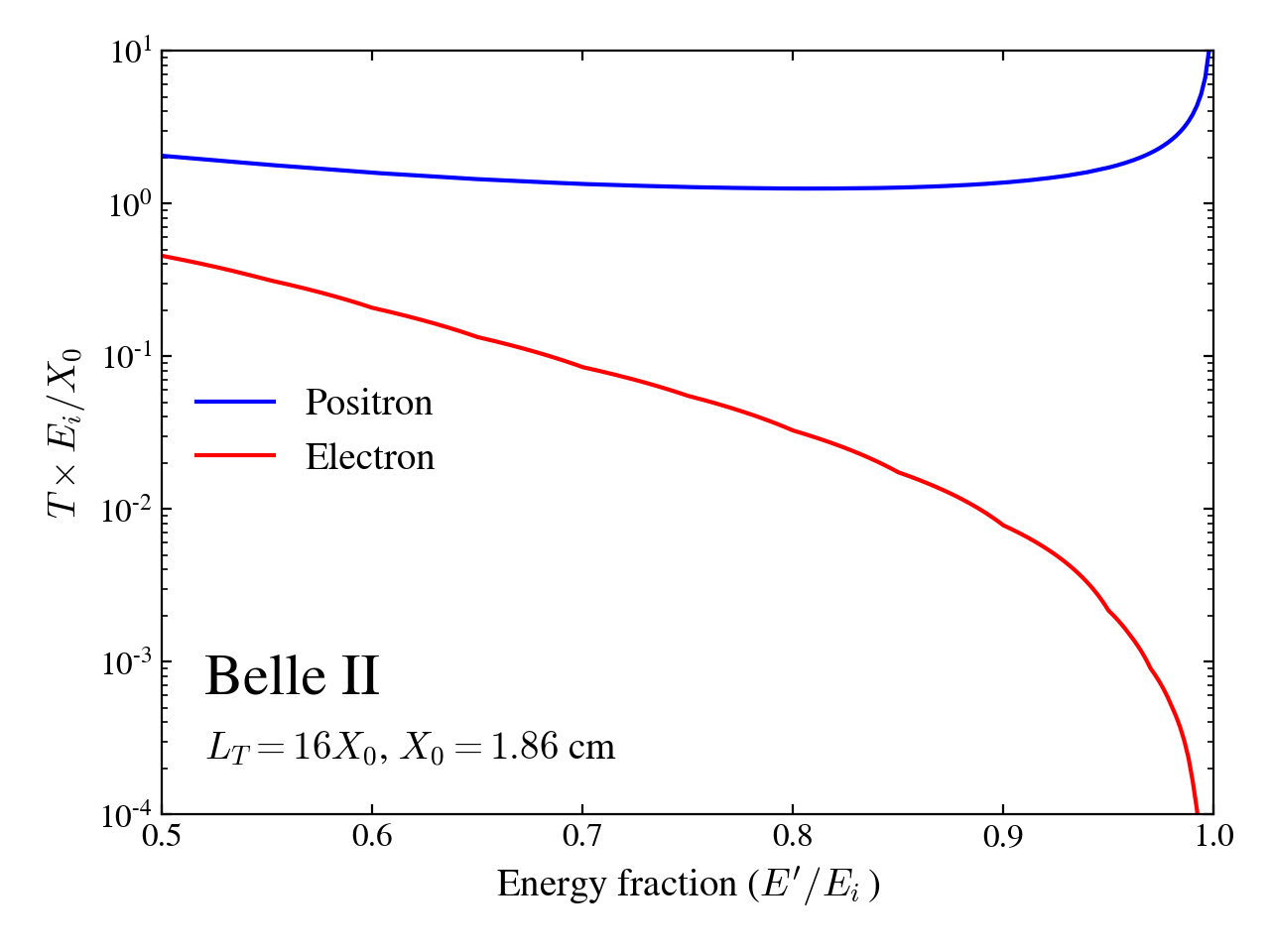} 
\caption{Differential track-length distribution $T$ normalized by $E_i/X_0$ for positrons (blue) and electrons (red) generated from the outgoing positrons of Bhabha scattering at Belle II.
$E'/E_i$ is the energy fraction carried by the incoming positron or electron,
$L_T$ is the target thickness and $X_0$ is the radiation length of the ECAL CsI crystal in Belle II.  
}
\label{fig:tract_length_dis}
\end{figure}

For an incident positron with energy $E_i$ that enters
a target with thickness $L_T$, the differential
track-length distribution of positions is a
function of the positron energy $E'$ \cite{Tsai:1966js},
\begin{align}
  T_e(E' , E_i , L_T)
=
  X_0 \int^{L_T/X_0}_0 dt \, I_e(E',E_i,t),
\end{align}
%
where $X_0$ is the radiation length of the target
and $I^{(i)}_e$, with $I_e \equiv \sum_{i} I^{(i)}_e$, denotes the
energy distribution of the $i$-th generation
positrons at the depth $tX_0$.
The formulas of
its first two generations are given by
\begin{align}
I^{(1)}_e (E',E_i,t) = \frac{1}{E_i} \frac{\left[ \ln (E_i /E') \right]^{\frac{4}{3}t -1}}{\Gamma \left( \frac{4}{3}t\right)},
\end{align}
\begin{align}
&
 I^{(2)}_e(E',E_i,t) =
  \int^t_0 dt' \int^{E_i}_{E'} dk
  \,
  I^{(1)}_e (E',k,t-t')
   \int^{E_i}_{k} dE_\gamma 
 \nonumber
 \\
 &
 \times
2I^{(1)}_\gamma (E_\gamma,E_i,t')
 \left[ \frac{4}{3}\left( 1- \frac{E_\gamma}{k} \right) + \left( \frac{E_\gamma}{k} \right)^2 \right] \frac{k^2}{E^3_\gamma},
\end{align}
where the energy distribution of the primary photon is 
\begin{align}
&
I^{(1)}_\gamma (E_\gamma,E_i,t') 
=
\frac{1}{E_\gamma} \int^{t'}_0 dt'' e^{-\frac{7}{9} ( t' - t'' )}
\int^{E_i}_{E_\gamma} dk'
\nonumber
\\ 
&
\quad
\times
I^{(1)}_e (k',E_i,t'')
 \left[ \frac{4}{3}\left( 1- \frac{E_\gamma}{k'} \right) + \left( \frac{E_\gamma}{k'} \right)^2 \right].
\end{align}
The differential track-length distribution of the secondary positrons for an incident electron at NA64 is formulated by
\begin{align}
\overline{T}_e(E' , E_i , L_T) = X_0 \int^{L_T/X_0}_0 dt \, I^{(2)}_e(E',E_i,t)~,
\end{align}
due to absence of the first generation positrons.
The above integral forms of the energy distributions
are still reliable at a thick target with $t > 1$,
such that the resulting track-length distributions
agree well with those in
\cite{Marsicano:2018krp,Marsicano:2018glj,NA64:2022rme}
obtained from GEANT4 simulations.
The electron and positron track-length distributions
are shown in \gfig{fig:tract_length_dis} for comparison.

\end{document}